\documentclass{appolb}
\usepackage{epsfig}
\preprint{FERMILAB-CONF-06-347-T}

\begin{document}
\title{The Standard Model: Alchemy and Astrology
\thanks{Presented at the conference ``Physics at LHC'', 3-8 July 2006, Polish
Academy of Arts and Sciences, Krakow.}%
}
\author{Joseph D. Lykken
\address{Fermi National Accelerator Laboratory\\
P.O Box 500, Batavia, IL 60510 USA}
}
\maketitle
\begin{abstract}
An brief unconventional review of Standard Model physics, 
containing no plots.
\end{abstract}
  
\section{Introduction}
In the days of Copernicus, the most prestigious activities at the Krakow
Academy were studies of alchemy and astrology. Since that time a number of
scientific revolutions, Copernican and otherwise, have advanced us to a
more sophisticated view of the universe based on particle physics and 
astrophysics. A Standard Model (SM) has emerged with a precise comprehensive
description of the constituents of matter and their interactions.
This model is, by far, the most predictive and best tested scientific
framework yet developed.

There is a temptation to regard the Standard Model as a fixed edifice,
completed and proofed. In the same spirit, the advent of the Large
Hadron Collider (LHC) is seen as the ground--breaking for a new framework
of physics beyond the Standard Model, with this new physics being the
dominant concern of the LHC program.

This way of thinking is incorrect. Our understanding of the Standard Model
has evolved greatly in the decades since the first elements of the
theory were put into place. This evolution will
continue until a number of profound mysteries are resolved. Discoveries
of physics beyond the SM may provide key insights, but the mysteries
themselves involve the structure and dynamics of the Standard Model proper.   

The LHC will indeed provide us with the first direct access to a new
framework of fundamental physics. The LHC will also probe the Standard 
Model in a new high energy, high luminosity environment. Both sorts of
activity will provide discoveries. They will also be tightly coupled,
since a better understanding of SM
processes will be required to extract the new framework, while the
new framework will elucidate mysteries of the Standard Model.
Far from marking the end of the Standard Model era, LHC turn--on
will induce a rash of technical and conceptual developments,
culminating in a much more sophisticated view of the same
constituents and interactions that comprise our current picture.

\section{The theoretical inputs of the Standard Model}

In order to understand the Standard Model, it is
enlightening to list the minimal set of theoretical inputs
that define it. Having done this, will we see in the next
section that the SM possesses a number of interesting derived
properties. This separation of inputs from derived properties
is ahistorical, \textit{i.e.}, I will employ modern insights
that were not available or not sufficiently appreciated at the time
that the SM was invented. 

\begin{table}[tb]
\centering
\begin{tabular}[b]{|c|c|c|c|}
\hline\hline 
label & $SU(3)_c$ & $SU(2)_L$ & $U(1)_Y$ \\ [0.5 ex] \hline
$Q_i$ & triplet & doublet & $1/3$ \\ [0.5 ex] \hline
$U_i$ & triplet & singlet & $4/3$ \\ [0.5 ex] \hline
$D_i$ & triplet & singlet & $-2/3$ \\ [0.5 ex] \hline
$L_i$ & singlet & doublet & $-1$ \\ [0.5 ex] \hline
$E_i$ & singlet & singlet & $-2$ \\ [0.5 ex] 
\hline\hline
\end{tabular}
\caption{The five varieties of fermions in the Standard Model.}
\end{table}

\begin{itemize}
\item{All interactions are local.}
\item{Quantum mechanics is correct, at least up to energy
scales around a TeV.}
\item{Special relativity, or more precisely four-dimensional
Poincar\'e invariance, is respected by interactions and
kinematics on these same scales.}
\end{itemize}
This set of assumptions implies that particles physics up
to some high energy scale
can be completely described by an effective relativistic
quantum field theory.

The next set of assumptions is about interactions:
\begin{itemize}
\item{There are gauge forces, mediated (at least at large
momentum transfers) by exchanges of gauge bosons.}
\item{The local gauge symmetry is $SU(3)_c\times
SU(2)_L\times U(1)_Y$.}
\item{Gravity exists but is ignored.}
\end{itemize}
The next set of assumptions is about constituents:
\begin{itemize}
\item{The fundamental matter constituents are two-component
complex Weyl fermions. They come in five varieties,
as shown in Table 1.}
\item{There are three copies (generations) of this
matter content.}
\end{itemize}
The next set of assumptions involve the Higgs:
\begin{itemize}
\item{the local gauge symmetry
$SU(3)_c\times SU(2)_L\times U(1)_Y$ is spontaneously
broken to $SU(3)_c\times U(1)_{em}$ via the Higgs mechanism.}
\item{This same Higgs scalar also has direct Yukawa
couplings to pairs of fermions. These pairs have the same
color charges but different hypercharges and weak charges:}
\begin{eqnarray}
\lambda^L_{ij}H\bar{L}_iE_j + \lambda^U_{ij}H^c\bar{Q}_iU_j + \lambda^D_{ij}
H\bar{Q}_iD_j + {\rm hermitian\ conjugates} \; .
\end{eqnarray}
\item{The matrices of Yukawa couplings are neither real nor diagonal.}
\end{itemize}
The final assumption looks rather obscure:
\begin{itemize}
\item{Only include operators up to dimension four.}
\end{itemize}
Historically, this assumption was included to make the theory
renormalizable. From a modern point of view, this 
input is poorly motivated.

\section{The derived properties of the Standard Model}

\subsection{mass and energy scales}
With one exception particle masses in the Standard Model are
derived from dynamics and dimensionless couplings. For example,
pure $SU(3)_c$ gauge theory is classically scale invariant,
but picks up a logarithmic scale dependence at one-loop
from the running of the gauge coupling. If I run this coupling
down from some arbitrary ultraviolet cutoff to the
infrared confining regime, I can trade the dimensionless gauge
coupling and the cutoff for a dimensionful dynamically determined
energy scale $\Lambda_{QCD}$.  

The only dimensionful input parameter of the Standard Model is the
negative Higgs mass-squared parameter $-m_H^2$. Together with the
dimensionless Higgs quartic self-coupling $\lambda$, these input
parameters determine the Higgs vacuum expectation value $v/\sqrt{2}$,
$v = 246.2$ GeV, as well as the Higgs mass. The derived value
of $v$, combined with the dimensionless SM Yukawa couplings, then
determine the masses of the quarks and leptons.

At finite temperature and vanishing chemical potential the SM has
two derived critical temperatures. One is the quark deconfinement
temperature of about 175 MeV, where we observe a crossover transition 
from hadrons to a quark--gluon plasma. The other is a temperature
of about 100 GeV where we expect a weakly first order phase transition
restoring the full $SU(2)_L\times U(1)_Y$ symmetry.

In the Standard Model neutrinos do not have mass, in contradiction
with experiment. I will return to this problem below.

\subsection{sowing the seeds of its own destruction}
Because we have forbidden higher dimension operators by hand,
the Standard Model has no explicit cutoff dependence. However,
if the Higgs self-coupling is too large -- corresponding to a physical
Higgs boson mass greater than about 180 GeV -- then the SM generates
its own ultraviolet cutoff $\Lambda_{LP}$. This is because $\lambda$
runs logarithmically with energy scale, and if $\lambda$ is large
enough at the electroweak scale the sign of the effect is to increase
$\lambda$ at higher energies. At some energy scale $\Lambda_{LP}$
the coupling hits a Landau pole and the electroweak sector
of the Standard Model breaks down.

If the Higgs self-coupling at the electroweak scale is too
small -- corresponding to a physical
Higgs boson mass less than about 130 GeV --
then the running goes the other way, and at some high energy scale
the sign of this quartic coupling goes negative. At best, this
destabilizes the vacuum; at worst, theories
with this kind of disease are unphysical. One could attempt to 
compensate by invoking dimension 6 Higgs self--couplings,
but this would violate one of our defining theoretical inputs.

\subsection{flavor}
The Standard Model has large accidental global flavor symmetries.
I call these accidental because if we had introduced generic higher dimension
operators \textit{ab initio}, these symmetries would be violated.
In any event most of them are not exact.

Baryon number $B$ and lepton number $L$ are accidental global symmetries.
They are exact at the perturbative level, but the combination
$B+L$ is broken by nonperturbative effects.

The SM has chiral symmetries for quarks and leptons, due to the
fact that gauge invariance forbids direct mass terms.
The chiral symmetries are broken nonperturbatively by a QCD
condensate, also breaking electroweak symmetry dynamically.
This dynamical effect, scaled up to much higher energies, is the basis
of technicolor models.

In the absence of Yukawa couplings, the Standard Model has a
huge $[U(3)]^5$ global flavor symmetry. These symmetries are
explicitly broken by the Yukawas, but since the Yukawas are
mostly quite small numerically, these symmetries are still
important.

Because the Yukawa matrices are not diagonal, the SM has flavor--changing
charged currents at tree level, as encoded in the CKM matrix.
The SM has no flavor--changing neutral currents (FCNCs) at tree level,
and at loop level FCNC's have an extra suppression (besides the loop
factor) coming from the GIM mechanism.

There are two sources of $CP$ violation in the Standard Model.
One is a single physical phase in the CKM matrix, coming from
the fact that the quark Yukawa matrices are complex. This phase
is rather large. The other
source of $CP$ violation is instantons, a nonperturbative effect
in QCD. This effect is parametrized by an angle $\theta_{QCD}$.
For unknown reasons this angle is either zero or very small,
$\theta_{QCD} < 10^{-9}$.

Last but not least, the SM has an accidental global $SU(2)$ symmetry
known as ``custodial $SU(2)$''. This is because the Higgs sector of the
SM has an $O(4)\sim SU(2)_L\times SU(2)_R$ global symmetry 
that acts on the four real components
of the complex Higgs doublet; custodial $SU(2)$ is the diagonal remnant
left unbroken after the Higgs gets a vev. This symmetry is broken
by the hypercharge gauge coupling and by the fermion doublet mass splittings.

\section{Disturbing features}

The Standard Model has many disturbing features. Some of these have
been nagging particle physicists for decades, while others have only
become apparent in recent years \cite{Lykken:2006gs}.

\subsection{the hierarchy problem}

The SM ignores gravity, which (modulo the possibility of extra
spatial dimensions) is an extremely good approximation for tree-level
processes in particle experiments. But the existence of gravity, combined
with naive (four-dimensional) scaling, implies
the existence of a Planckian regime $M_{\rm Planck} \sim 10^{19}$ GeV
where gravity becomes strong. There are presumably new Planckian 
degrees of freedom associated with this threshold. In the absence of
supersymmetry, the SM Higgs should interact with these states via
loops. So why isn't $|m_H| \sim M_{\rm Planck}$?

Suppose that this problem is somehow solved. Then we observe that
the SM has a number of potential gauge anomalies. These anomalies
want to induce a one-loop breaking of the local $SU(2)_L\times U(1)_Y$ 
gauge invariance. The only reason that this does not happen is that
the fermion content of each generation exactly matches the 15
SM nonsinglet members of the 16 of $SO(10)$ (the sixteenth state
can be added to provide neutrino masses).
Furthermore, if we extrapolate the SM gauge couplings to higher energies,
we find that they roughly unify at a scale $M_{\rm GUT} \sim 10^{14}$ GeV
(the unification is more precise, though still not perfect, if we
add the assumption of a supersymmetry threshold 
at $\sim 1$ TeV \cite{Chung:2003fi}).
Thus we have two strong hints that the SM has an underlying
grand unified structure. So why isn't $|m_H| \sim M_{\rm GUT}$?

Suppose that this problem is somehow solved. Then we go back to our
previous observation that the SM sows the seeds of its own destruction,
through the running of the Higgs self-coupling $\lambda$. Over 95\%
of the allowed mass range for the Higgs, this implies a mass scale
$M_{LP}$ or $M_{VI}$ at which the SM breaks down due to a Landau pole
or a vacuum instability. So why isn't $|m_H| \sim M_{\rm LP}$?

\subsection{flavor}

In the Standard Model the quark and lepton Yukawa couplings are
inputs. Since the couplings run, their precise values are dependent
on both scale and renormalization scheme; further subtleties arise in
extracting the Yukawas of the light quarks. But roughly speaking, if these
couplings were true input parameters, determined \textit{e.g.} by
initial conditions of the early universe, we would expect them
to be of order one. 

Instead the SM Yukawas are hierarchical, with values as low as
$\sim 3 \times 10^{-6}$, and intrafamily mass ratios as large as 40.
The Yukawa mixings also have a hierarchical structure, as evidenced
by the famous Wolfenstein parametrization of the CKM matrix:
\begin{eqnarray}
V_{CKM} = \pmatrix{1-\lambda^2/2 & \lambda & A\lambda^3
(\rho - i\eta ) \cr -\lambda & 1-\lambda^2/2 & A\lambda^2\cr
A\lambda^3(1-\rho -i\eta ) & -A\lambda^2 & 1} + {\cal O}(\lambda^4)
\; ,
\end{eqnarray} 
where $\lambda$ is a small parameter ($\lambda \simeq 0.2$).

There is only one diagonal Yukawa coupling that is of order one,
and that is the top quark Yukawa. But even this case is mysterious.
The top Yukawa is not really {\it of order} one: it is equal to one!
For example, using the 2005 combined Tevatron value for the pole mass
of the top quark, the corresponding Yukawa coupling is
$\lambda_t = 0.99 \pm 0.01$. The entire particle physics community
has chosen (so far) to regard this fact as a 1\% coincidence.
I should point out that similar percent level equalities \textit{e.g.}
supersymmetric gauge coupling unification or the ratio of the
total mass-energy density of the universe to the critical density,
have spawned huge theoretical frameworks bolstered by thousands of papers.

Obviously the flavor structure of the Standard Model is {\it not} random,
and is begging for explanation. It is still possible for some
of this hierarchical
structure to arise from initial conditions in the early universe,
if we invoke anthropic arguments and allow ourselves the decadent
luxury of positing $10^{500}$ different vacuum bubbles out beyond
the Hubble horizon. But the more straightforward and economical
explanations are (i) the Standard Model is formulated with too 
few degrees of freedom:
the Yukawa couplings should be promoted to fields, whose vacuum expectation
values are determined by a combination of dynamics and symmetries, or
(ii) the Standard Model is formulated with too many degress of freedom:
the quarks and leptons are not fundamental, and the flavor structure
is a feature of the dynamics that maps the true fundamental constituents
(strings, preons, etc) to the light SM states that we observe.

\subsection{higher dimension operators}

From the discussion above of the hierarchy problem it seems impossible
that the Standard Model lagrangian provides an accurate description
of nature at arbitrarily high energies. Thus we should regard the SM
as an effective field theory. We then expect that we have probably neglected 
higher dimension operators constructed out of SM fields and suppressed
by powers of an ultraviolet cutoff $\Lambda > v$. Of course there may
be several different cutoffs involved, representing a variety of physical
thesholds. In the early days of the SM it made
sense to neglect such operators, since our experiments for the most part
probed energy scales less than $v$.

In more recent times many experiments have probed SM processes at
energy scales much larger than $v$, either directly (at the Tevatron),
through processes sensitive to loops (at LEP, the B factories, etc)
or through processes sensitive to new small violations of the accidental
symmetries of the Standard Model.
Dozens of experiments have had the opportunity to observe
the effects of dimension 5 and dimension 6 operators constructed
out of SM fields, including both operators that violate accidental
symmetries of the SM and operators that preserve those symmetries.  

With one important exception, none of these experiments have observed
clear evidence for any such higher dimension operators. This surprising
result is worth reviewing in some detail. To organize out thinking,
I will divide the SM higher dimension operators into three classes:
those that violate $B$ and/or $L$, those
that violate the approximate flavor symmetries of the
SM, introducing new sources of flavor violation besides those already
provided by the SM Yukawa matrices, and
those that respect all SM accidental symmetries, \textit{i.e.}
respect $B$ and $L$ and are Minimal Flavor Violating (MFV).

\subsubsection{$B$ and/or $L$ violating operators}

Experimental bounds on proton decay and on charged lepton flavor
violating processes ($\mu\to e\gamma$, $\tau\to \mu\gamma$, $\mu\to e$
conversion) tell us that generic dimension 6 operators that violate
$B$ and $L$ either do not exist or are suppressed by a superheavy mass
scale. This is an important result, indicating that conservation
of $B$ and $L$ is not so accidental after all.

The discovery of neutrino masses means that the SM requires some
kind of extension. By adding right-handed neutrinos (SM singlet
fermions) as new degrees of freedom, we can preserve the coupling
rules for the SM. An alternative that does not require new degrees
of freedom is to introduce the unique (gauge invariant) dimension 5
operator that can be constructed out of SM fields:
\begin{eqnarray}
{\cal O}_5 = {f\over\Lambda}
\left( L^TCi\tau_2\vec{\tau}L\right)
\left(H^T i\tau_2\vec{\tau}H\right) \; ,
\end{eqnarray}
where the $\tau_i$ are $SU(2)_L$ matrices, and $f$ is a dimensionless
coupling with generation indices suppressed.
This operator violates lepton number by 2 units, and gives Majorana
masses to neutrinos $m_{\nu} \sim fv^2/\Lambda$.
If we require $f$ of order one for the heaviest neutrino
and take the heaviest neutrino to saturate the current experimental
upper bound of a few tenths of an eV, then the cutoff scale
$\Lambda$ is a few times $10^{14}$ GeV. The Majorana nature of the
neutrino mass may be confirmed in the near future by the observation
of neutrinoless double beta decay.

Thus it appears that neutrino data favors augmenting the SM by its unique
dimension 5 operator, but this operator is suppressed by
a superheavy mass scale. It is not at all clear what this implies
about the likelihood of observing any of the large number of dimension
6 operators. 

\subsubsection{flavor violating operators}

Many dimension 6 operators would provide new sources of quark flavor
violation beyond that induced by the CKM matrix. A large number
of experiments have been performed in the $b$, charm and kaon sectors
looking for such effects. So far no clear signals have been observed
anywhere, and impressive limits have been set. Some FCNC operators
are only compatible with experiment if they are suppressed by
a cutoff exceeding 1000 TeV.

This situation may change with the next round of experiments,
but currently the simplest interpretation of this data is
that the CKM matrix is the only source of quark flavor violation,
up to scales of a few TeV or higher.

\subsubsection{symmetry preserving operators}

An important class of dimension 6 operators are the ``oblique''
operators. These operators are purely electroweak, flavor
diagonal, and at leading order they only affect the $W$ and $Z$ vacuum
polarization. They violate no accidental symmmetries except for
custodial $SU(2)$. If present, these operators would shift the values of the
oblique parameters $S$ and $T$ \cite{Peskin:1991sw}.
The latest global fits to electroweak precision data show
no clear evidence for any such effects \cite{lepewwg}.
Higgs naturalness implies that such operators are
likely to exist, suppressed by a cutoff that
is no larger than a couple of TeV, \textit{i.e.} $|m_H|$ divided
by the square root of a loop factor. The current experimental
lower bounds exceed this estimate.  

There is also a large class of dimension 6 four-fermion
operators that are flavor diagonal. There is no
argument based on SM symmetries that would forbid such operators. 
Nevertheless
they are not seen in data. The cutoff scale for such operators
is constrained, \textit{e.g.} to be larger than a 10 TeV for $eeee$ couplings
and 26 TeV for $eedd$ couplings \cite{Yao:2006px}.

\section{Discovering the Standard Model}

By the 1980s the basic elements of the Standard Model were clearly
defined, and many key predictions had been spectacularly verified by
experiment. However two particles predicted by the Standard Model -- the top
quark and the Higgs boson -- had not been observed. Not only did the
SM predict the existence of these particles, it also predicted the values
of all of their quantum numbers, except their masses.

Standard Model radiative corrections contain diagrams with virtual
top quarks and Higgs bosons. This leads to electroweak observables
whose predicted values depend logarithmically on the ratios
$m_t^2/m_Z^2$ and $m_h^2/m_Z^2$, where $m_h$ is the mass of the
physical Higgs particle. Note we could replace $m_Z$ with $m_W$ in
these ratios, since the difference is higher order.
There are also leading order corrections that are directly
proportional to $m_t^2$, rather than to a logarithm. These arise from
the Yukawa couplings of the top to the Goldstone bosons that were
eaten by the $W^{\pm}$ and $Z$ gauge bosons, and indicate the fact
that top radiative corrections do not decouple in the limit
$m_t \to \infty$ with $m_b$ fixed.
With the advent of the LEP experiments
and SLD, it was possible to observe the quadratic $m_t^2$ effects
of virtual tops in precision electroweak data. This was a big discovery.

\subsection{the Tevatron top}
In 1995, the Tevatron experiments discovered a new strongly pair-produced
state that decays promptly to a $W$ boson and a $b$-jet. This was
a big discovery. Of course such a state is compatible with the top
quark predicted by the SM, and its mass was compatible with the
less precise mass determinations from electroweak precision data.

During this past year, data from Run II of the Tevatron has
allowed us for the first time to probe many properties of this
new heavy particle, comparing this Tevatron top with the theoretical
particle of the Standard Model. Here is a quick summary of what has been
discovered:
\begin{itemize}
\item{The Tevatron top has charge 2/3. A measurement of the jet
charges of the $b$-jets produced from $t\bar{t}$ pairs now
eliminates the possibility of a charge 4/3 particle at nearly
95\% confidence \cite{Dzero:2006vd}.}
\item{ The Tevatron top has spin 1/2, a coupling to the $W$ that
is more like $V-A$ than $V+A$, and a coupling to longitudinal $W$'s
consistent with the SM Higgs mechanism to within an experimental
uncertainty of about 20\% . These results come from measuring the
angles between the charged $e$ or $\mu$ and the $b$-jet in top
decays. The distribution of these angles allows a fit to
$f_0$ and $f_+$, the fraction of decays that produce a longitudinally
polarized $W$ and a right-handed $W$, respectively. If the Tevatron
top had spin 3/2, $f_+$ would be close to 1; if it had spin 1/2
but a $V+A$ coupling, $f_+$ would be close to 0.3. For a SM top,
producing a right-handed $W$ requires a $b$ quark helicity flip, suppressing
this decay by the ratio $m_b^2/m_t^2$. Results from 
CDF and DZero \cite{CDFtophel,Abazov:2006hb}
show $f_+ < 0.09$ at 95\% confidence, and $f_0$ within about
20\% of its SM value.}
\end{itemize}

These top results are just an example of many recent discoveries
in Standard Model physics. One effect of these discoveries
has been to rule out or place tight constraints on scenarios
beyond the Standard Model, but the broader significance is that
we are observing for the first time what Nature is really doing in these
fundamental phenomena of particle physics.

\subsection{the virtual virtual Higgs}

With the LHC (and perhaps even the Tevatron) we expect a
similar story to unfold regarding a new particle (or particles)
that I will generically call the Higgs. It is often said,
invoking the famous ``blue band'' plot,
that the electroweak precision data already
reveals the radiative effects of a light Higgs, much as the
effects of virtual top were observed before the Tevatron discovery.

Strictly speaking, this claim is false, as can be seen by
have a closer look at the global electroweak fits. For example,
let me use the analysis of Appendix E of a recent combined
analysis \cite{LEP:2005em}.
The data is used to fit the purely electroweak 
quasi-observables \cite{Altarelli:1997et}
$\epsilon_1$, $\epsilon_2$, $\epsilon_3$ and $\epsilon_b$.
The effects of virtual top are clearly seen in these fits;
\textit{e.g.} using the leading order relation
\begin{eqnarray}
m_t^2 = -4\pi\epsilon_b{\sqrt{2}\over G_F}
\end{eqnarray}
I obtain the (unsophisticated) estimate $m_t = 155\pm 25$ GeV.
Virtual SM Higgs effects can arise to leading order from
two different combinations:
\begin{eqnarray}\label{eqn:epsfits}
3\epsilon_3 -\epsilon_2 \propto {\rm ln}\,{m_h\over m_Z}
&=& -0.002 \pm 0.004 \; ,\\
2\epsilon_1 + 3\epsilon_b  \propto {\rm ln}\,{m_h\over m_Z}
&=& -0.004 \pm 0.005 \; ,
\end{eqnarray}
where my error bars do not take into account correlations.

Thus we have two way of detecting a virtual Higgs in existing
data, and in both cases we have obtained a null result,
despite part per mil accuracy. The prevailing {\it interpretation}
of these results (with which I agree) is that the Higgs
is light, with a mass not much above $m_Z$ in logarithmic units.
The reason why this interpretation is reasonable is because
if there were no Higgs we would have expected to see some {\it other}
radiative effects in the electroweak data.

Thus the statement that the precision data favors a light Higgs,
as opposed to no Higgs at all, relies upon some theoretical baggage.
This baggage originates from the observation that SM diagrams
for longitudinal $WW$ and $WZ$ scattering give amplitudes that
grow like (energy)$^2$ and (energy)$^4$, violating unitarity
at energies a little above a TeV \cite{Lee:1977eg}.
Adding the SM Higgs restores
a weakly coupled theory. Other alternatives have been explored,
in which Kaluza-Klein gauge bosons or new strong interactions
do the job of Higgs in restoring unitarity. Generically such 
alternatives do produce fairly large radiative effects, but no
one claims to know that this is necessarily true in all cases.

\subsection{QCD}

We have ample evidence that QCD is the correct theory of
strong interactions, and QCD has an unambiguous nonperturbative
definition via Wilsonian ideas applied on the lattice.
Nevertheless many fundamental questions about QCD are still
unanswered \cite{Ellis:2006ev}. 
One way to gauge our ignorance is to ask for a
detailed picture of the interior of a proton. We know that the
answer to this question depends on the nature of the process
used as a probe, in particular on the squared momentum transfer
$Q^2$ and the parton momentum fraction $x$. We know that there
are \textit{at least} three qualitatively different regimes: large $x$ +
large $Q^2$, large $x$ + small $Q^2$, and small $x$ + small $Q^2$.
The first is the standard perturbative regime, and is the only one
in which we have a detailed picture of proton constituents. Yet
even here we are mystified by basic issues such as how the spin
of the proton is distributed among the partons. For the second regime
we have a vague picture of valence quarks confined by gluonic flux
tubes, but no detailed understanding. In the third regime, which
is quite relevant for LHC, we are just beginning to grapple with
the dynamics of parton saturation.

\subsection{prediction}

During the LHC era, we will discover as many important new insights
about the Standard Model as we discover about physics beyond the
Standard Model. A decade from now, we will look back on our current
understanding of the Standard Model and be amused at its lack of
sophistication. 

\section*{Acknowledgments}
Fermilab is operated
by Universities Research Association Inc. under Contract
No. DE-AC02-76CHO3000 with the U.S. Department of Energy.

\end{document}